\documentclass[aps,pre,twocolumn,showpacs,superscriptaddress]{revtex4}
\usepackage{amssymb,amsmath,amsthm}
\usepackage[dvips]{graphicx,color,epsfig}
\begin{document}

\title{Complexity, information transfer and collective behavior \\ in chaotic dynamical networks}
\author{M. Escalona-Mor\'an} 
\affiliation{Departamento de C\'alculo, Escuela B\'asica, Facultad de Ingenier\'ia, Universidad de Los Andes, 
M\'erida, Venezuela. }
\author{G. Paredes}
\affiliation{Laboratorio de F\'isica Aplicada y Computacional, \\Universidad Nacional Experimental del T\'achira, 
San Crist\'obal, Venezuela.}
\author{M. G. Cosenza}
\affiliation{Centro de F\'isica Fundamental, Universidad de Los Andes, M\'erida, Venezuela.}

\begin{abstract}
We investigate the relationship between complexity, information transfer and the emergence of collective behaviors, such as  synchronization and nontrivial collective behavior, in a network of globally coupled chaotic maps
as a simple model of a complex system.  We calculate various quantities for this system: the mean field, a measure of statistical complexity, the information transfer, as well as the information shared, between the macroscopic and local levels as functions of
the strength of a coupling parameter in the system. Our results show that the emergence of nontrivial collective behavior is associated to higher values of complexity. Little transference of information from the global to the local level occurs 
when the system settles into nontrivial collective behavior while no information at all flows between these two scales in a synchronized collective state. As the parameter values for the onset of nontrivial collective behavior or chaos synchronization are approached, the information transfer from the macroscopic level to the local level is higher, in comparison to the situation where those collective states are already established in the system. Our results add support to the view of complexity as an emergent collective property that is absent at the local level in systems of interacting elements.
\end{abstract}

\maketitle

\section{Introduction}
In recent times, the study of complex systems has become a new paradigm for the search of a unified description of
the  mechanisms for emergence of organization, structures and functionality in natural and artificial phenomena 
in diverse scenarios (Mikhailov and Calenbuhr, 2002; Kaneko and  Tsuda, 2000; Boccara, 2004).
One common feature found among the prevalent viewpoints on the meaning of complexity is emergent
behavior: collective structures, patterns and functions that are absent 
at the local level arise from simple interaction rules between the constitutive elements in a system.
Phenomena such as the formation of spatial patterns, collective
oscillations, synchronization of chaos in spatiotemporal systems, spiral waves, segregation and differentiation,
growth of domains, motion of swarms, bird flocks, and fish schools, opinion consensus, and economic crashes are examples
of self-organizing processes that occur in various contexts such as
physical, chemical, physiological, biological and social and economic systems.

The phenomenon of synchronization, where all the state variables of a system converge to a single trajectory in phase space, constitutes one of the simplest and more common collective behaviors occurring in networks of mutually interacting dynamical elements (Pecora and Carroll, 1990; Pikovsky et al., 2002; Manrubia et al., 2004). In particular, the study of chaos synchronization has provided insights into many natural processes and practical
applications such as secure communications and control of dynamical systems (Boccaletti et al., 2002; Uchida  et al., 2005; Argyris et al., 2005).

On the other hand, nontrivial collective behavior can also emerge in
systems of interacting chaotic elements (Chat\'e and Manneville, 1992). 
This phenomenon is characterized by an ordered evolution of macroscopic quantities  coexisting with 
local chaos. Synchronization and other collective behaviors in globally coupled oscillators are relevant in many chemical and biological systems and have been experimentally investigated (Wang et al. 2000; De Monte  et al., 2007; Taylor et al., 2009). Models based on coupled map networks have
been widely used in the investigation of collective phenomena
that appear in many complex systems ( Kaneko and  Tsuda, 2000). In particular, nontrivial collective behavior occurs in networks of coupled chaotic maps, such as regular Euclidean lattices (Chat\'e and Manneville, 1992), one-dimensional lattices (Cosenza, 1995), fractal geometries (Cosenza, 1998), and globally coupled systems (Kaneko, 1990; Shibata et al., 1999).

In this paper, we investigate the relationship between complexity and the emergence of collective behaviors, such as  synchronization and nontrivial collective behavior, in chaotic dynamical network models. We investigate the information
transfer, as well as the information shared,  between the global and local levels of the network as a condition for
complexity or self-organization in spatiotemporal systems. Specifically, we address the questions: how much
information does a local unit possess about the collective dynamics of a system? or how do the information flow
and the complexity depend on parameters of a system?.

In Section 2, we present a globally coupled chaotic map network as a simple example of a system of interacting nonlinear elements that shows emergent collective behaviors. We calculate a measure of statistical complexity, introduced by L\'opez-Ruiz et al. (L\'opez-Ruiz et al., 1995), as a function of a parameter of the system and show that the appearance of nontrivial collective behavior is associated to higher values of complexity in comparison to those values for synchronization or for disordered states. Similarly, we show that, as the parameter values for the onset of nontrivial collective behavior or chaos synchronization are approached, the information transfer from the macroscopic level to the local level is higher, in comparison to the situation where those collective states are already established in the system. Our results are discussed in the Conclusions. 

\section{Characterization of the collective behavior of a network of coupled chaotic maps.}
We consider a system of $N$ globally coupled chaotic maps, where the state of map $i$ ($i=1,2,\dots,N$) at discrete time $t$ is denoted by $x_t^i$. The evolution of the state of each map is assumed to depend on its own local dynamics and on its interaction with all the other maps in the system. Thus, we consider a network of maps subjected to a mean field global interaction (Kaneko, 1990)
\begin{equation}
\label{systemEq}
x_{t+1}^i = (1-\varepsilon)f_i (x_t^i)+ \frac{\varepsilon}{N} \sum_{j=1}^N f(x_t^j),
\end{equation}
where the function $f(x_t^i)$ describes the local dynamics of element $i$ and $\varepsilon$ is a parameter expressing the strength of the coupling. As local chaotic dynamics we choose a map belonging to the family of singular maps $f(x_t)=b-| x_t |^z$,
where $|z| < 1$ and $b$ is a real parameter (Alvarez-Llamoza et al., 2008). These maps do not belong to the standard class of universality of unimodal or bounded maps. They exhibit robust chaos, with no periodic windows in a finite interval of the parameter $b$ that depends on $z$, for  $|z| < 1$. Robustness is an important property in applications that require reliable operation under chaos in the sense that the chaotic behavior cannot be destroyed by small perturbations of the system parameters.
In this paper we employ the value $z=-0.25$ for which the corresponding singular map displays
robust chaotic dynamics in the range $b \in [0.9896,1.6493]$ (Alvarez-Llamoza et al., 2008).

We shall use the following macroscopic quantities to characterize the emergence of collective behavior in the system  Eq.~\ref{systemEq},

1) The instantaneous mean field,
\begin{equation}
\label{meanFieldEq}
S_t = \frac{1}{N}\sum_{j=1}^N f(x_t^j).
\end{equation}

2) The asymptotic time-average $\langle\sigma\rangle$ of the instantaneous standard deviations $\sigma_t$ of the distribution of map variables $x^i_t$, defined as
\begin{equation}
\langle\sigma\rangle = \frac{1}{T}\sum_{t=\tau}^{\tau+T}  \sigma_t,
\end{equation}
\begin{equation}
\sigma_t=\left[ \frac{1}{N} \sum_{i=1}^N \left( x^i_t - \frac{1}{N}\sum_{j=1}^N x_t^j \right)^2 \right]^{1/2},
\end{equation}
where $\tau$ is a number of discarded transients. A completely synchronized state in the system Eq.~\ref{systemEq} occurs when $x_t^i=x_t^j$, $\forall i,j$.
Stable synchronization corresponds to $\langle \sigma \rangle=0$. 

3) The statistical complexity, defined as (L\'opez-Ruiz et al., 1995)
\begin{equation}\label{complexityEq}
C = H \cdot D = -K\sum_{s=1}^{R} p_s \log p_s \cdot \sum_{s=1}^{R} \Big(p_s - \frac{1}{R}\Big)^2 \, ,
\end{equation}
where $H$ is the entropy and $D$ is the disequilibrium, a sort of \textit{distance} to the equipartition in a system; 
$p_s$ represents the probability associated to the state $s$;  $R$ is
the number of states that the system possesses at the given level of description, 
and $K$ is a positive normalization constant. Note that $p_s$ may vary for different
levels of observation, reflected in $R$. The measure $C$ has been shown to  be capable of discerning among different 
macroscopic structures emerging in complex systems at a given scale (S\'anchez and  L\'opez-Ruiz, 2005). Thus, according to this point of view, the level of complexity of a system is given by the interplay between
the entropy and the disequilibrium. For the system Eq.~\ref{systemEq}, we shall consider the probability distribution of the states $x_t^i$. 

4) The information transfer from the dynamical variable $y_t$ to the variable $x_t$ in an interacting system, is defined as
(Schreiber, 2000)
\begin{equation}
T_{y \rightarrow x}= \sum_{x_{t+1},x_{t},y_{t}} p\left( x_{t+1},x_{t},y_{t}\right) log\left( \frac{p\left( x_{t+1},x_{t},y_{t}\right)p\left( x_{t}\right) }{p\left( x_{t},y_{t}\right)  p\left( x_{t+1},x_{t}\right) } \right) ,
\end{equation}
where $p(x_t)$ means the probability distribution of the time series $x_t$, $p(x_t,y_t)$ is the joint probability distribution 
of $x_t$ and $y_t$, and so on. The quantity $T_{y \rightarrow x}$ measures the degree of dependence of $x$ on $y$; i.e., the information required to represent the value $x_{t+1}$ from the knowledge of $y_t$. Note that the information transfer is nonsymmetrical, i.e., $T_{y \rightarrow x} \neq T_{x \rightarrow y}$.

5) The mutual information shared by two subsystems $y$ and $x$ is defined as (Shannon and Weaver, 1949) 
\begin{equation}
M_{x,y}= \sum_{x_t,y_t} p\left( x_t,y_t\right) log\left( \frac{p\left( x_t,y_t\right) }{p\left( x_t\right) p\left( y_t\right) }\right) .
\end{equation}
The quantity $M_{x,y}$ measures the overlap of the information content of the variables $x$ and $y$; it represents how much the uncertainty about $x$ decreases if $y$ is known. The mutual information $M_{x,y}$ is symmetrical and does not indicate the direction of the flow of information between two interacting dynamical variables, as $T$ does. 

We consider the coupled map network, Eq.~\ref{systemEq}, for a system of size  $N=10^5$. The local singular maps have exponent $z=-0.25$ and their parameter is fixed at $b=1.1$, within the robust chaos regime.  Figure~1 shows the above macroscopic variables
calculated as a function of the coupling parameter $\varepsilon$ for the system  Eq.~\ref{systemEq}.

Figure~1(a) shows the bifurcation diagram of a map $x_t^i$ in this system as a function of $\varepsilon$. For each value of $\varepsilon$, the value of $x_t^i$ is  plotted at each time step during a run of $10^3$ iterates starting from random initial conditions on the local maps, uniformly distributed on the interval $x_0^i\in[-8,2]$, after discarding $10^3$ transients. At the local level, the dynamics is chaotic over the entire range of $\varepsilon$. 

Figure~1(b) shows the bifurcation diagram of the mean field $S_t$ for the system, Eq.~\ref{systemEq}, as a function of $\varepsilon$. For the same initial conditions  and discarded transients as in Fig.~1(a), $10^3$ consecutive values of $S_t$ were calculated for each value of $\varepsilon$. The mean field in Fig.~1(b) reveals the presence of global periodic attractors
for some range of the coupling parameter. Different collective states
emerge as a function of the coupling $\varepsilon$: a turbulent phase (T),
where $S_t$ manifests itself as a global fixed point, a state where the time series of mean field fluctuates around a single value;
collective periodic states coexisting with local chaos, corresponding to nontrivial collective behavior (NTCB); collective chaotic bands (C); and chaotic
synchronization (S). In this representation, collective periodic
states at a given value of the coupling appear as sets
of vertical segments which correspond to intrinsic fluctuations
of the periodic oscillations of the mean field. At a value $\varepsilon=0.04$, a pitchfork bifurcation takes place from a statistical fixed point to a collective period-two state, where the time series of $S_t$ alternatingly moves between the corresponding neighborhoods of two separated, well-defined values. Increasing the coupling induces the emergence of collective states of higher periodicity. Global attractors of period $2$, $4$, and $8$ are visible in Figure~1(b).
On the other hand, increasing the
system size $N$ does not decrease the amplitude of the collective
periodic orbits. Moreover, when $N$ is increased the
widths of the segments that make a periodic orbit in the
bifurcation diagrams such as in Fig.~1(b) shrink, indicating
that the global periodic attractors become better defined in
the large system limit. This phenomenon of nontrivial collective behavior is an example of emergent behavior in a complex dynamical system. 

Figure~1(c) shows the quantity $\langle\sigma\rangle$ versus $\varepsilon$ for the system Eq.~\ref{systemEq}, averaged over a run of $T=10^3$ iterates after discarding $\tau=10^4$ transient for each value of $\varepsilon$. The turbulent phase (T) corresponds to the state of lower coherence, manifested by the higher value of $\langle\sigma\rangle$ in this region of the coupling parameter. There is a critical value $\varepsilon=0.32$ at which $\langle\sigma\rangle$ drops to zero, indicating that the chaotic elements in the system become synchronized (S). 

Figure~1(d) shows the statistical complexity $C$ of the mean field as a function of $\varepsilon$. Here, the observation level was set at $R=64\times 10^3$. When the value of $\varepsilon$ is small, the mean field of the system follows the standard
statistical behavior of uncorrelated disordered variables that is reflected in the single period in the bifurcation diagram of  $S_t$. At the chosen level of resolution, the complexity measure considers 
the macroscopical variable $S_t$ as laying in a single state, thus giving a small value of $C$ in the region T. 
The complexity $C$ remains small up to 
a critical value of the coupling $\varepsilon \simeq 0.04$, where $C$ suddenly increases, resembling a phase transition. 
As the periodicity of the collective orbit 
increases, more states are occupied by the probability distribution of the mean field $S_t$.
The probability distribution of $S_t$ corresponding to 
a periodic collective state is not uniform and consists of a set of distinct ``humps''.  A nonuniform probability distribution and few occupied states lead to larger values of the complexity $C$, as observed in the region NTCB.
When the system enters chaotic collective band motion (C), more states are occupied by the probability distribution of the mean field and therefore this probability becomes more uniform. As a consequence, $C$ decreases. In the region of chaotic synchronization (S), the complexity is low. There, $S_t=f(x_t^i), \,  \forall i$, and therefore macroscopic behavior can be 
trivially derived from the local behavior, yielding low complexity for the system, as one may expect. 
On the other hand, Figure~1(d) shows that maximum complexity is associated to the emergence 
of nontrivial collective behavior. The occurrence of ordered collective behavior in the coupled map network, Eq.~(\ref{systemEq}),
cannot be attributed to the existence of windows of periodicity in the local dynamics.

Figure~1(e) shows the information transfer $T_{S \rightarrow x^i}$ from the mean field $S_t$ to one element $x_t^i$ in the system, Eq.~(\ref{systemEq}), as a function of $\varepsilon$. The number of states used to calculate the probability distributions is $600$. We observe that the information transfer is moderate in the turbulent region, where the local states are uncorrelated. In the region of nontrivial collective behavior, the flow of information from the mean field to a local map decreases; the evolution of global and the local variables become correlated, 
reflecting the appearance of self-organization in the system. The information transfer  $T_{S \rightarrow x^i}$ increases in the region of chaotic bands behavior and reaches a maximum value just before the onset of synchronization. To achieve chaotic synchronization, the flow of information from the global level to the local levels should be large. Once chaotic synchronization is reached in the S region, the behavior of $S_t$ is identical to that of the maps; they do not longer depend on mean field  signal. As a consequence,  $T_{S \rightarrow x^i}$ vanishes in this region. 

Finally, Figure~1(f) shows the mutual information $M_{S,x^i}$ between the mean field $S_t$ and one element $x_t^i$ in the system, Eq.~(\ref{systemEq}), as a function of $\varepsilon$. In the turbulent and NTCB regions, there is little similarity shared by the global and local variables in the system and $M_{S,x^i}$ is small. However, as the coupling $\varepsilon$ increases, the variables become more correlated and the mutual information reaches a maximum plateau in the synchronization region, as one may expect when  variables become identical. 

\section{Conclusions}
We have calculated several quantities to characterize the emergence of collective behavior in
a system of coupled chaotic maps as a function of a parameter expressing the strength of the coupling between the maps. 
We have employed maps displaying robust chaos as local dynamics because the emergence of ordered collective behavior in this
kind of dynamical networks cannot be attributed to the existence of windows of periodicity at the local level, but to the interactions between the constitutive elements. 

The mean field contains relevant information about the collective behavior of the system.
Our results show that the emergence of nontrivial collective behavior in spatiotemporal systems is associated to higher values of complexity in comparison to those values for synchronization and for disordered states. The increase of complexity can be interpreted as a manifestation of collective organization in the system. 

Connectivity and coupling strengths are the mechanism for information
exchange in networks of dynamical units. 
We have found that little transference of information from the macroscopic to the local level occurs 
when the system settles into nontrivial collective behavior and that no information is transferred at all between these two scales in a synchronized collective state. However, the information shared by the maps in the system is maximum at synchronization, as one may expect. These results can be related to those of Cisneros et al. (Cisneros et al., 2002) 
who showed that the prediction error used to measure the mutual prediction error between a local and a 
global variable in a network of chaotic maps decreased when nontrivial collective behavior arises in the system.

The information transfer that is required for the appearance of
nontrivial collective behavior and chaotic synchronization takes place at some specific
values of the parameters of the system. As the critical parameter values for the onset of nontrivial collective behavior or chaos synchronization are approached, the information transfer from the macroscopic level to the local level is greater, in comparison to the situation where those collective states are already established in the system.  This result suggests that a surge of information flow from the global to local variables may serve as a predictor in parameter space for the occurrence of coherent or collective behavior in complex systems.

Nontrivial collective behavior is a global property of the system 
that is neither present at the local level nor induced externally. 
Our results add support to the view of complexity as an emergent behavior
in systems of interacting elements. Morever, our results indicate that the statistical complexity and the information transfer can be useful quantities to characterize the transitions to various types of collective behaviors in dynamical chaotic networks and in other complex systems of interacting elements. 

\section*{Acknowledgment}
This work was supported by Consejo de Desarrollo Cient\'{\i}fico, Human\'{\i}stico 
y Tecnol\'ogico, Universidad de Los Andes, Venezuela, under grant C-1694-10-05-B.

\section*{References} 
 \begin{description}

\item Alvarez-Llamoza, O., Cosenza, M. G. and Ponce, G. A. 2008. Critical behavior of the Lyapunov exponent
in type-III intermittency. \textit{Chaos, Solitons and Fractals} \textbf{36}: 150-156.

\item  Argyris, A., Syvridis, D.,  Larger, L., Annovazzi-Lodi, V.,  Colet, P., Fischer, I., 
Garcia-Ojalvo, J., Mirasso, C. R., Pesquera, L. and Shore, K. A. 2005.  Chaos-based communications at high bit rates using commercial fibre-optic links. {\it Nature (London)} {\bf 438}: 343-346.

\item  Boccaletti, S.,  Kurths, J., Osipov, G., Valladares, D. L., \&  Zhou, C. S. 2002. The synchronization of chaotic systems. {\it Phys. Rep.} {\bf 366}: 1-101. 

\item Boccara, N. 2004. \textit{Modeling complex systems}, Springer-Verlag, New York. 

\item Chat\'e, H. and Manneville, P.  1992. Emergence of effective low-dimensional dynamics in the macroscopic behaviour of coupled map lattices, \textit{Europhys. Lett.} {\bf 17}: 291-296. 

\item Cisneros, L., Jim\'enez, J.,  Cosenza, M. G. and 
Parravano, A. 2002. Information transfer and nontrivial collective behavior in chaotic coupled map networks,  \textit{Phys. Rev. E} {\bf 65}: 045204R. 

\item Cosenza, M. G. 1995. Collective behavior of coupled chaotic maps, \textit{Phys. Lett A} \textbf{204}: 128-132.

\item Cosenza, M. G. 1998. Nontrivial collective behavior in coupled maps on fractal lattices, \textit{Physica A} {\bf 257}: 357-364. 

\item  De Monte, S., d'Ovidio, F., Dan\o{}, S., S\o{}rensen, P. G. 2007. Dynamical quorum sensing: population density encoded in cellular dynamics. {\it Proc. Natl. Acad. Sci.} {\bf 104}: 18377-18381.  

\item Kaneko, K. 1990. Clustering, coding, switching, hierarchical ordering, and control in networks of chaotic elements, \textit{Physica} \textbf{D 41}: 137-172.

\item Kaneko, K. and Tsuda, I. 2000. \textit{Complex Systems: Chaos and beyond}, Springer, Berlin. 

\item L\'opez-Ruiz, R.,  Mancini, H. L. and  Calbet, X. 1995. A statistical measure of complexity, \textit{Phys. Lett. 
A} {\bf 209}: 321-326. 

\item Manrubia, S. C.,  Mikhailov, A. \& Zanette, D. H. 2004. \textit{Emergence of Dynamical Order}, 
World Scientifc, Singapore.

\item Mikhailov, A. and Calenbuhr, V. 2002. \textit{From Swarms to Societies: Models of complex behavior}, 
Springer, Berlin.

\item Pecora, L. M. and Carroll, T. L. 1990. Synchronization in chaotic systems, \textit{Phys. Rev. Lett.} \textbf{64}: 821-824.

\item Pikovsky,  A., Rosenblum,  M.  and Kurths, J. 2002. \textit{Synchronization: a
universal concept in nonlinear sciences}, Cambridge University Press, Cambridge.

\item S\'anchez, J. R. and  L\'opez-Ruiz, R. 2005. A method to discern complexity in two-dimensional patterns
generated by coupled map lattices, \textit{Physica A} {\bf 355}: 633-640.

\item Schreiber, T. 2000. Measuring information transfer, {\it Phys. Rev. Lett.} {\bf 85}: 461-464. 

\item Shannon, C. E. and Weaver, W. 1949. \textit{The Mathematical Theory
of Information}, University of Illinois Press, Urbana, IL.

\item  Shibata, T., Chawanya, T., and Kaneko, K. 1999. Noiseless collective motion out of noisy chaos, {\it Phys. Rev. Lett.} {\bf 82}: 4424-4427. 

\item  Taylor, A. F., Tinsley, M. R.,  Wang, F., Huang, Z. and Showalter, K. 2009. Dynamical quorum sensing and synchronization in large populations of chemical oscillators, {\it Science} {\bf 323}: 614-617.  

\item  Uchida, A., Rogister, F., Garcia-Ojalvo, J. and Roy,  R. 2005.  Synchronization and communication with chaotic laser systems. {\it Prog. Opt.} {\bf 48}: 203-341. 

\item  Wang, W., Kiss, I. Z.,  and Hudson, J. L. 
 2000. Experiments on arrays of globally coupled chaotic electrochemical oscillators: synchronization and clustering,
{\it Chaos} {\bf 10}: 248-256. 

\end{description}

\begin{figure}[h]
\begin{center}
\includegraphics[scale=0.6]{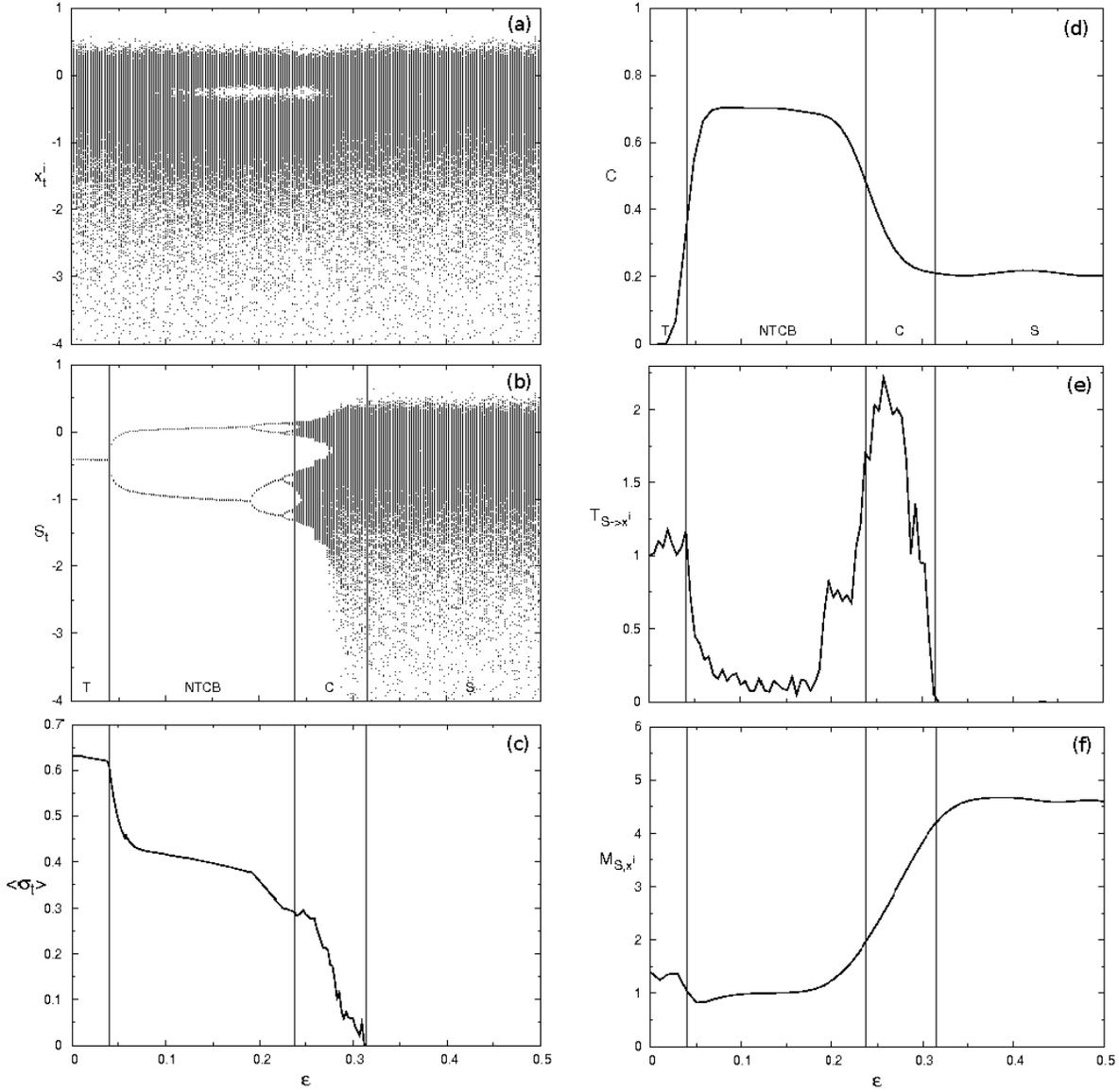} 
\caption{Various quantities to characterize the collective behavior of the system Eq.~\ref{systemEq} as a function of the coupling $\varepsilon$. Fixed local parameter is $b=1.1$; local singularity exponent $z=-0.25$; system size is $N=10^5$. Random initial conditions on the local maps, uniformly distributed on the interval $x_0^i\in[-8,2]$ are used. (a) Bifurcation diagram of a map $x_t^i$   versus $\varepsilon$. (b) Bifurcation diagram of the mean field $S_t$ versus $\varepsilon$. The labels indicate T: turbulent, NTCB: nontrivial collective behavior, C: collective chaotic bands, S: chaotic synchronization. (c) $\langle\sigma\rangle$ versus $\varepsilon$. (d) Statistical complexity $C$ of the mean field versus $\varepsilon$. (e) Information transfer $T_{S \rightarrow x^i}$ 
versus $\varepsilon$, averaged over $10$ maps. (f) Mutual information $M_{S,x^i}$ versus $\varepsilon$.}
\end{center}
\end{figure}

\end{document}